\documentclass[conference]{IEEEtran}
\IEEEoverridecommandlockouts

\usepackage[utf8]{inputenc}
\usepackage[T1]{fontenc}

\usepackage{booktabs}
\usepackage{subcaption}
\usepackage[tstt]{backnaur}

\usepackage[frozencache,cachedir=.]{minted}

\usepackage{amsmath}
\usepackage{hyperref}
\usepackage{cleveref}
\usepackage{amsfonts}
\usepackage{mwe}
\usepackage{prftree}
\usepackage{listings}
\usepackage{algorithm}
\usepackage{algpseudocode}

\usepackage{xcolor}
\usepackage{siunitx}
\usepackage{multirow}
\usepackage{makecell}
\usepackage{tabularx}
\usepackage{mwe}
\usepackage[inline]{enumitem}

\DeclareMathAlphabet{\mathcal}{OMS}{cmsy}{m}{n}
\SetMathAlphabet{\mathcal}{bold}{OMS}{cmsy}{b}{n}

\begin{document}
\title{
  Type-Directed Program Synthesis and Constraint Generation for Library Portability
  \thanks{This work was supported by the Engineering and Physical Sciences
  Research Council (grant EP/L01503X/1), EPSRC Centre for Doctoral Training in
Pervasive Parallelism at the University of Edinburgh, School of Informatics.}
}

\author{
  \IEEEauthorblockN{Bruce Collie}
  \IEEEauthorblockA{
    \textit{School of Informatics} \\
    \textit{University of Edinburgh}\\
    \textit{United Kingdom} \\
    bruce.collie@ed.ac.uk}
  \and
  \IEEEauthorblockN{Philip Ginsbach}
  \IEEEauthorblockA{
    \textit{School of Informatics} \\
    \textit{University of Edinburgh}\\
    \textit{United Kingdom} \\
    philip.ginsbach@ed.ac.uk}
  \and
  \IEEEauthorblockN{Michael F.P.\ O'Boyle}
  \IEEEauthorblockA{
    \textit{School of Informatics} \\
    \textit{University of Edinburgh}\\
    \textit{United Kingdom} \\
    mob@inf.ed.ac.uk}
}

\maketitle

\begin{abstract}
Fast numerical libraries have been a cornerstone of scientific computing
for decades, but this comes at a price. Programs may be tied to vendor
specific software ecosystems resulting in polluted, non-portable code.
As we enter an era of heterogeneous computing, there is an explosion in
the number of accelerator libraries required to harness specialized
hardware. We need a system that allows developers to exploit
ever-changing accelerator libraries, without over-specializing their
code.

As we cannot know the behavior of future libraries ahead of time, this
paper develops a scheme that assists developers in matching their code
to new libraries, without requiring the source code for these libraries.

Furthermore, it can recover equivalent code from programs that use
existing libraries and automatically port them to new interfaces. It
first uses program synthesis to determine the meaning of a library, then
maps the synthesized description into generalized constraints which are
used to search the program for replacement opportunities to present to
the developer.

We applied this approach to existing large applications from the
scientific computing and deep learning domains. Using our approach, we
show speedups ranging from 1.1$\times$ to over 10$\times$ on end to end
performance when using accelerator libraries.

\end{abstract}

\maketitle


\section{Introduction}
\label{sec:intro}
Fast numerical libraries have been a cornerstone of scientific computing for
decades \cite{Sedova,Pfluger:2016:STS:3019106.3019110}.  They provide efficient
implementations of key algorithmic components and allow a separation of
concerns.  This comes at a cost, however, as it may  tie programs into
vendor-specific software ecosystems and results in non-portable, polluted code.
A new library API means that the original ``vanilla'' code has to be recovered,
and then modified to use the new libraries. 

The risk of being tied into an out of date library API has led some developers
to release multiple versions of their code, e.g.\ PyTorch \cite{pytorch} and
Darknet \cite{darknet13}. This requires the maintenance of multiple code bases
and complex build systems. However, as we witness the rise of specialized
heterogeneous accelerators, we also see a proliferation of accelerator libraries
\cite{cublas,clblas,cudnn,cldnn}. In the long-term, a multi-versioned code base
is not sustainable.

This paper develops an alternative approach: a compiler-based scheme that
discovers opportunities to use new accelerator libraries in user code, with
little prior knowledge of the libraries. Furthermore, it can recover
behaviorally equivalent code from programs that use existing libraries and
automatically port them to new interfaces. In order to reduce developer burden,
it attempts to do this with minimal intervention using program synthesis and
graph matching.

Program synthesis is a well studied area that deals with searching a program
space to find candidate programs that match a specification \cite{Lee2018}.  Our
program synthesis implementation uses   a number of generic control-flow
components and a set of  heuristics  defining when they should be applied.
These heuristics  are driven by a library's type signature  and lightweight
annotations provided by the library vendor. Crucially, these annotations are
easily extracted from documentation and  require no knowledge of a library's
internals.

Once we know what a library does, we need to see if the developer's program has
structures that match its behavior. We achieve this by automatically describing
the synthesized program as a set of constraints which we then use to search the
user code. As the synthesized program may not easily match existing code, we
first generate many equivalent versions, normalize and then generalize them to a
common description that determines the most appropriate constraints. When we
find a match we suggest to the developer that a replacement could be made to
take advantage of a different library.

Our synthesis and constraint generation allows us to target large existing code
bases and show significant performance improvement.  We applied this approach to
existing large applications from scientific computing and deep learning written
in C, C++ and Fortran. We show speedups ranging from 1.1$\times$ to over
10$\times$ improvement when implementing replacements suggested by our tools.

\section{Example}
\label{sec:motivation}
\begin{figure*}[t]
  \centering
  \includegraphics[width=\textwidth]{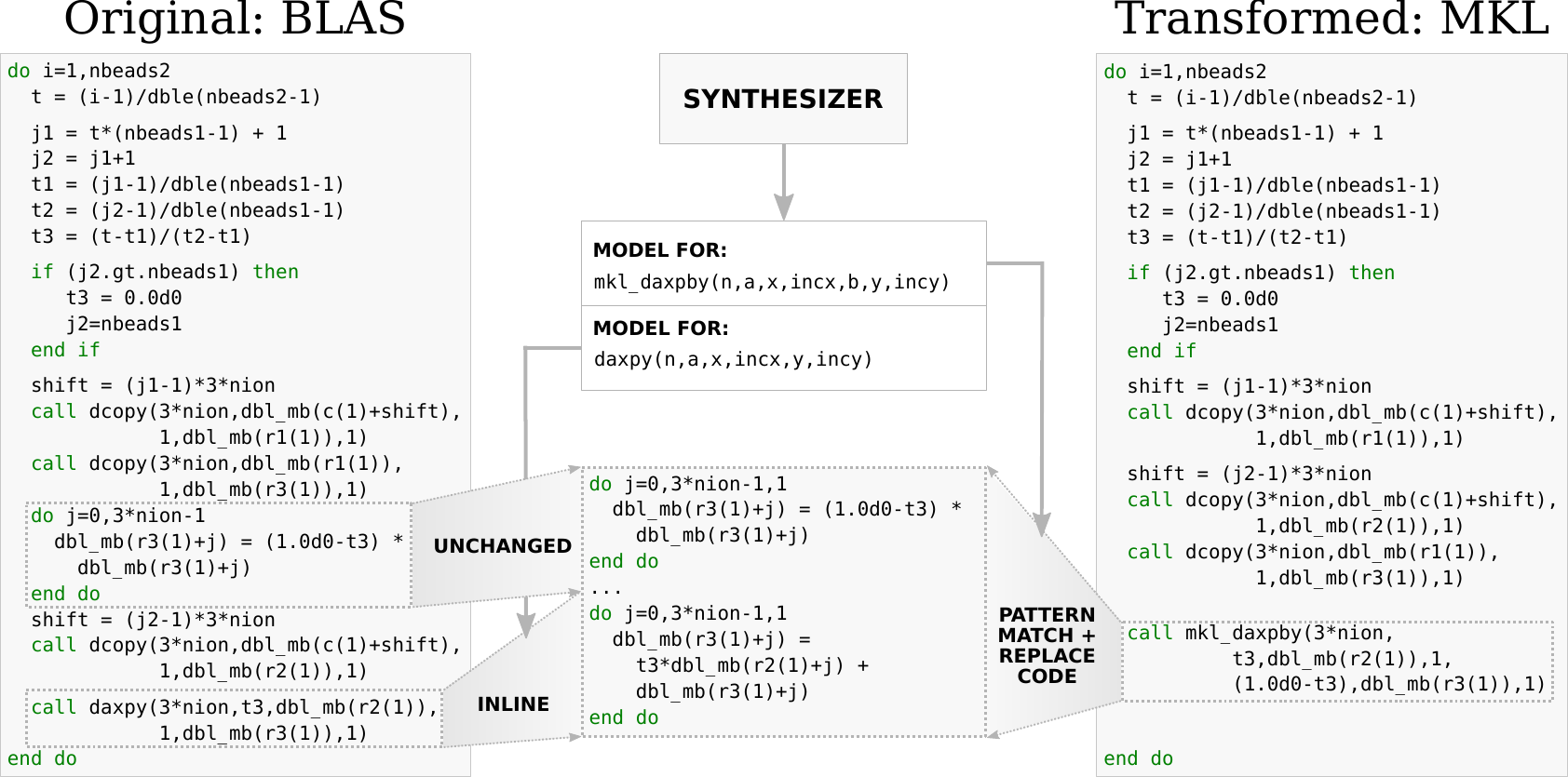}
  \caption{Porting BLAS to MKL, managing API evolution. 
  On the left is code taken from NWChem \cite{Valiev2010}, a widely-used
  chemical simulation suite.
  Our approach learns and inlines the code for {\tt daxpy},
  then identifies the resulting code as being equivalent to the model
  learned for {\tt mkl\_daxpby}.}
  \label{fig:motivation}
\end{figure*}

This section illustrates how our approach helps in porting code that
uses an existing library API to a new library API with increased
functionality. Consider the code sample on the left of
\Cref{fig:motivation}. This is an inner loop taken from a subroutine in
NWChem \cite{Valiev2010}, a widely used chemical simulation suite that
makes explicit calls to BLAS libraries. The code contains manual loops
over arrays and calls to BLAS routines ({\tt dcopy} and {\tt daxpy}). We
wish to port this code to use Intel's MKL libraries (as shown on the
right hand side, which makes use of the extended MKL functionality {\tt
mkl\_daxpby}).

On the left hand side of the figure, there are two highlighted sections of code.
The first highlighted piece of code is a simple loop that performs the
following abstract vector operation:
\begin{equation}
  \mathbf{r_3}  \gets (1 - t_3) \mathbf{r_3} 
\end{equation}
The second highlighted piece of code is a call to {\tt daxpy}.
{\em If} we had access to the {\tt daxpy} source code, we would see this
corresponds to the following  vector operation:
\begin{equation}
  \mathbf{r_3}  \gets t_3 \mathbf{r_2} + \mathbf{r_3}
\end{equation}
As we wish to port NWChem to MKL, we can exploit the  extended MKL
\cite{mkl} BLAS function {\tt mkl\_daxpby}. This supports vector
scaling, combined with {\tt daxpy} corresponding to the vector operation
\begin{equation}
  \mathbf{r} \gets  (1 - t) \mathbf{r} +t \mathbf{r'} 
\end{equation}
A call to this extended implementation
{\tt mkl\_daxpby} is shown in the highlighted box on the RHS of
\Cref{fig:motivation}.
The two original operations in equations (1,2) can be rewritten as a single
operation by substituting for $\mathbf{r_3}$ in equation (2) to give 
\begin{equation}
  \mathbf{r_3} \gets t_3 \mathbf{r_2} + (1 - t_3) \mathbf{r_3}
\end{equation}
which corresponds to the extended function {\tt daxpby}. In this
example, it means that we can legally replace the two highlighted pieces
of code on the LHS of the figure with the similarly highlighted code on the RHS.

{\bf Match and Replace:} If there is a source level description of both
versions of {\tt daxpy}, we first inline the original call, as shown in
the {\bf Inline} box in \Cref{fig:motivation} to give the modified code
in the central part of the figure.\footnote{Calls to other functions
(e.g. \texttt{dcopy}) will also be inlined, but are not shown in the
figure for clarity.} We then try to match this modified code to the code
corresponding to the extended version from MKL \cite{mkl}. We achieve
this using a novel graph based constraint solver which matches the two
codes and suggests to the developer  that the old calls be replaced with
the new one.

{\bf Synthesis:} In practice, we cannot guarantee there is a suitable
source level description of every use of a library. This may be due to
the library provider not releasing an appropriate description; it no
longer being available; or being poorly documented. It may also be
defined in a manner suitable for human consumption but not compiler
automation. It is certainly the case that there is not agreement among
all library developers about a universal language to describe the
semantics of all their libraries.  

\begin{figure*}[t]
  \centering
  \includegraphics[width=\textwidth]{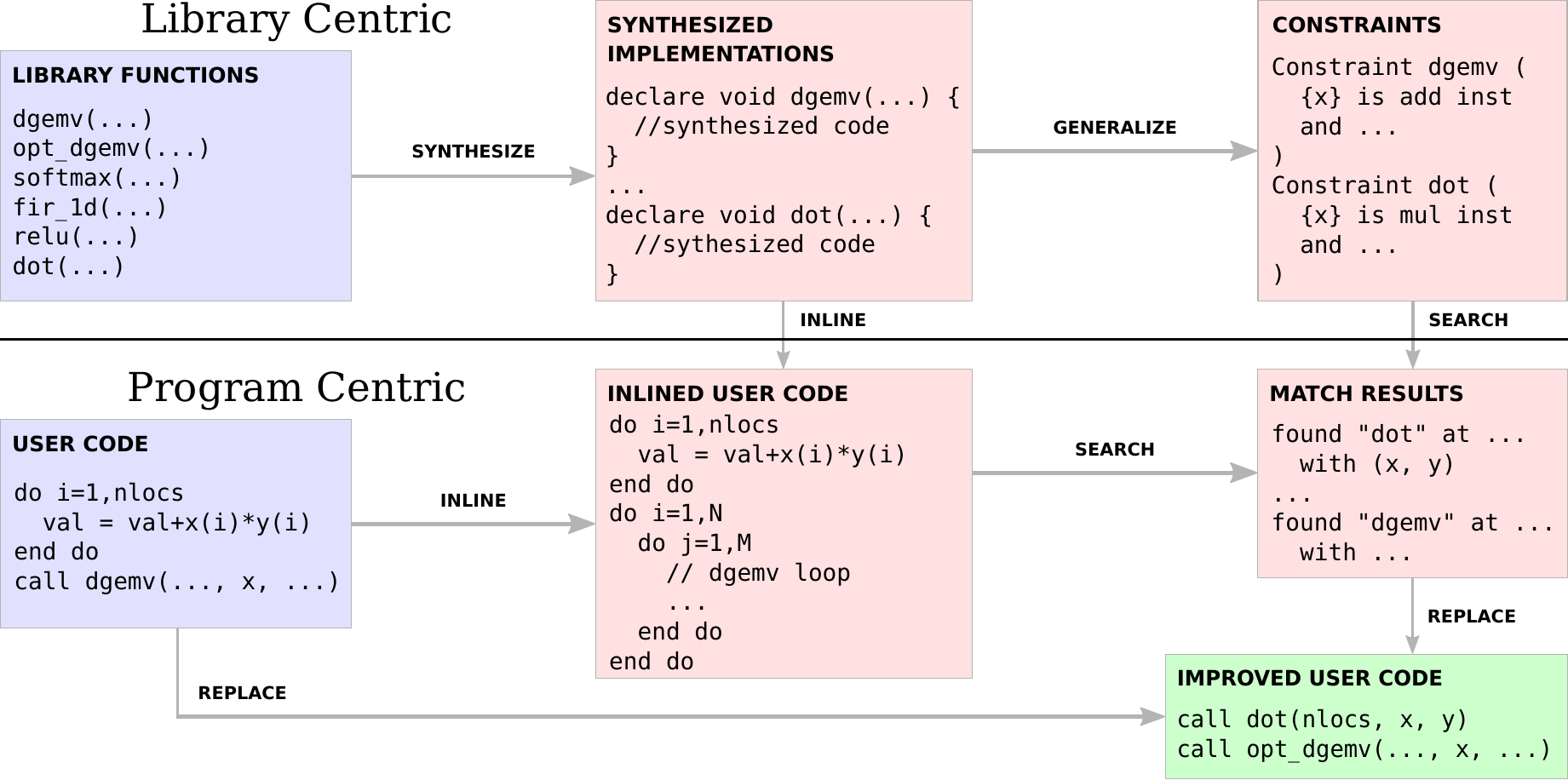}
  \caption{An overview of the data flow through the different stages of our
  system.
  Inputs are shown in blue on the left, intermediate products in red and the
  final product in green in the bottom right corner.
  For brevity, the content shown in each box is an approximation of the actual
  data in the system.}
  \label{fig:overview}
\end{figure*}
 
If the source code of the two libraries is unavailable, we use program
synthesis to generate programs that correspond to both versions of {\tt
daxpy} as shown by the {\bf Synthesizer} box in \Cref{fig:motivation}.
We then, as before, inline the original BLAS {\tt daxpy}, pattern match
and suggest to the developer that it be replaced with a call to the
extended MKL version.

Thus, the developer  is able to port their code to a new, extended
library without having to identify the opportunity manually (they need
only agree to the suggested replacement). In this example, it results in
a 20\% performance improvement on an Intel Xeon E5-2620. If the code is
to be ported again or an improved library is released, then the
procedure can be repeated, avoiding legacy API tie-in. At the heart of
our approach is the use of program synthesis and graph based generalized
constraint matching.

\section{Overview}
\label{sec:overview}
\Cref{fig:overview} gives a high level overview of our approach. It can
be split into two sections: {\em library centric} and {\em program
centric}. The library centric work corresponds to the top three boxes in
the diagram and has to be performed once for each new library
considered. The program centric work corresponds to the remaining boxes
and is performed when a user program has to be re-targeted to a new API.

\subsection{Library Centric}

Given the extended type of a library, we use oracle guided program synthesis
\cite{Jha2017} to generate a program that is equivalent to the behavior of the
library. This is achieved by generating many input/output pairs which are used
to guide synthesis. As the space of programs is unbounded, we exploit type
information and heuristics to generate synthesized programs in a reasonable
time.

Once we have a candidate synthesized implementation, it is unlikely to
have an identical structure to all user code. We therefore generate many
versions and then generalize and derive a constraint program that
describes the generalized program structure.  These constraints will be
used later to search the user programs for equivalent matches.  As we
work within the LLVM compiler infrastructure, we synthesize programs and
derive constraints at the LLVM static single assignment intermediate
representation of a program.

\subsection{Program Centric}

Once a user program is to be ported, we first inline the synthesized
implementation of any library calls. If there is a source code
description of the library available, this can be used, otherwise we
rely on the synthesized code.  Next we search for code patterns
corresponding to the constraints derived from the library centric phase. 

\subsection{Contribution}

The program centric phase is based on prior work and uses the constraint
language IDL, the SMT solver and replacement technique described in
\cite{Ginsbach2018}. Our contribution is restricted to the library centric
phases where we make two novel contributions:
\begin{itemize}
  \item Oracle-guided LLVM program synthesis of black-box imperative
    libraries with nested control structures using lightweight type
    signature annotations.
  \item Automatic generation and generalization of program constraints
    from examples using graph matching.
\end{itemize}
We elaborate on these in \Cref{sec:method} and \Cref{sec:conversion}
before evaluating our approach on a range of real-world applications.

\subsection{Practical Usage}

Neither the synthesis nor the generalization procedures in
\Cref{fig:overview} can guarantee semantic correctness; instead we rely
on the notion of behavioral equivalence as described in
\Cref{sec:behavioral}. In practice, we rely on the developer to sign off
on any code replacement. To avoid wasting developer time with false
positives, each potential replacement can be first checked by comparing
the output of the original code against the suggested replacement. Only
successful candidates are then presented.  \Cref{sec:results} describes
the ways in which both the library and program centric components may
exhibit unsoundness, and show that in practice the usefulness of our
tools is not greatly affected.

\section{Learning Programs}
\label{sec:method}
\subsection{Annotated Signatures} \label{sec:annotated}

Automatically learning the behavior of an arbitrary function, given only
its type signature, is generally an intractable problem. We define a
simple language of annotations (in the spirit of a minimal logic
programming language) that can be used by library vendors to annotate
their functions with arbitrary additional semantic information not
expressible in a type signature.

\begin{figure}[t]
  \centering
  \includegraphics[width=0.84\columnwidth]{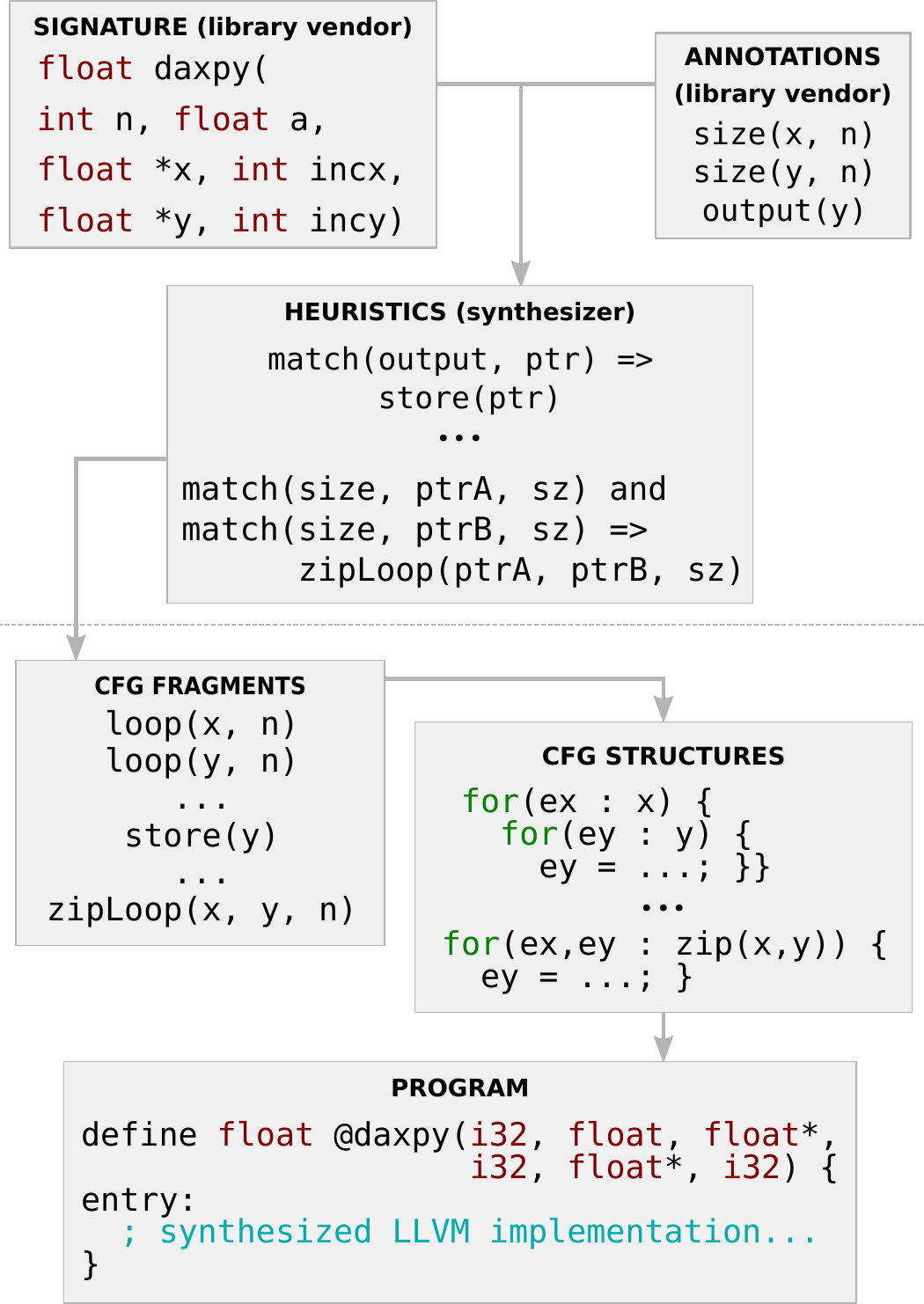}
  \caption{A simplified illustration of how our synthesizer learns an
    implementation for the \texttt{daxpy} linear algebra function. Inputs to the
    process (from the library vendor and the synthesizer itself) are given above
    the dotted line, and the synthesis process is given below. The set of
  annotations given (\texttt{size} and \texttt{output}) is complete, but the
heuristics, fragments and instantiated CFG compositions have been abbreviated
for brevity. A full explanation of this worked example is given in
\Cref{sec:heuristics}}
  \label{fig:synth}
\end{figure}

A useful motivating example is encoding the relationship between a
pointer to allocated memory and the size of that memory. The upper right
corner of \Cref{fig:synth} shows this being used to annotate the
function \texttt{daxpy} from \Cref{fig:motivation}. A full listing of
the annotations used in this paper is given in
\Cref{sec:annotations}---the properties corresponding to these
annotations are conceptually simple and can be easily extracted from API
documentation.

\subsection{Annotation Details} \label{sec:annotations}

The descriptions and algorithms given above are abstract and could be
used to instantiate many different program synthesizers, depending on
the annotations and fragment templates used. The functions evaluated in
this paper are synthesized using five core annotations, each of which is
listed and briefly specified below:

\begin{description}
  \item[\texttt{size(xs, n)}:] the pointer \texttt{xs} points to allocated
    memory with \texttt{n} elements.
  \item[\texttt{output(x)}:] the pointer \texttt{x} is an output parameter for
    the function.
  \item[\texttt{enum(x, c0, ..., cN)}:] the parameter \texttt{x} must take one
    of the distinct constant values \texttt{c0}$\dots$\texttt{cN}.
  \item[\texttt{pack(xs, c)}:] each logical entry in the array pointed to by
    \texttt{xs} contains \texttt{c} physical elements.
  \item[\texttt{indices(xs)}:] elements of \texttt{xs} in memory are logically
    array indices.
\end{description}

Almost all of the functions we synthesize use at least one \texttt{size}
annotation. The use of separate pointer and size arguments is endemic to
C function signatures, and we found documentation highlighting this
relationship in every library considered. Similarly, logical output
arguments are always highlighted in documentation (some degree of
inference could be performed for this annotation by considering
\mintinline{C}{const}-specified pointers). The annotation \texttt{enum}
is used for BLAS functions that perform (for example) transposed or
non-transposed versions of the same computation, and we found the values
were listed prominently in documentation.

The remaining two annotations are less closely tied to documentation,
but were used in only a small number of cases. \texttt{pack} was used to
simplify implementation details for functions dealing with homegeneous
structure types, and all uses of it could easily be removed with no
conceptual changes. Only one use of \texttt{indices} was
necessary---when synthesizing \texttt{spmv}, to ensure that indirect
memory accesses were safe. This annotation could be found in
documentation, albeit less prominently than \texttt{size} and
\texttt{output}.

These annotations were sufficient to synthesize all the functions
described in \Cref{sec:results}. We do not believe supplying them
represents a significant burden on the library vendor, and could easily
be automated in the simpler cases.

\subsection{Control Flow} \label{sec:heuristics}

Our approach to synthesizing candidate programs is two-phase. Firstly, we
construct and sample from the set of potential control-flow structures that a
candidate might use. Secondly, we fill in the control structures with
instructions.

\paragraph{Control Flow Fragments} 
A control flow fragment is a region of code with `holes' in it, such
that they can be composed with other fragments to form a complete
program CFG. The holes in a fragment may be filled by any other
fragment, but may also remain empty; a valid LLVM IR programs can be
extracted from any composition. Additionally, fragments may be
parameterized.

\paragraph{Querying Properties} 
In order to generate potentially valid control flow structures, we
require a set of candidate fragments that might comprise a solution.
This part of the synthesis process is driven by inference rules
expressing heuristics on when each type of control flow fragment should
be instantiated. For example, the first line in the \textsc{heuristics}
box in \Cref{fig:synth} shows a rule for instantiating a loop over
memory if the size of that memory is known.  Queries use a limited form
of unification with conjunctive matches (e.g.\ \texttt{sz} in
\Cref{fig:synth} when instantiating a \texttt{zipLoop}).

\begin{algorithm}[t]
  \caption{Dataflow generation}
  \label{alg:dataflow}
  \begin{algorithmic}[1]
    \Function{FillDataflow}{$cfg, n$}
      \State $ tree \gets $ dominance tree of $ cfg $
      \State $ phis \gets \emptyset $
      \For{each $block$ in $ \mathrm{inorder}(tree) $}
        \If{$block$ has $ > 1 $ predecessors}
          \State $ phis \gets phis \; \cup $ \{ new $ \phi $ node in $block$ \}
        \EndIf
        \If{$block$ is a dataflow block}
          \State $ live \gets $ live SSA values in $ block $
          \State generate $ n $ instructions sampling from $live$
        \EndIf
      \EndFor
      \For{each $ \phi $ in $ phis $}
        \State $ live \gets $ live SSA values at block with $\phi$
        \State choose incoming values to $\phi$ from $live$
      \EndFor
    \EndFunction
  \end{algorithmic}
\end{algorithm}

\paragraph{Sampling Control Flow}
The first step in the synthesis process is to enumerate all the possible
query matches against the function type signature and property
annotations. This yields a set of control flow fragments, from which the
synthesizer will construct program fragments by composition. To do this,
we perform an exhaustive search over the possible compositions of up to
3 fragments. If no solution is found with 3 fragments, we revert to a
random sampling process, although we have not found this to be necessary
in practice.

\paragraph{Example} 
\Cref{fig:synth} shows a worked example of this synthesis procedure for
the \texttt{daxpy} function used in \Cref{sec:motivation}. The top two
boxes show its type signature and the full set of easily-obtainable
library vendor annotations required for synthesis. These annotations are
matched against the synthesizer's set of heuristics, shown in the
central box. Only the two most important ones are shown: creating a
store instruction if a pointer is a logical output, and iterating over
pointers with the same logical size together.  Then, the synthesizer
collects the full set of possible fragments; some less useful ones such
as individual loops over \texttt{x} and \texttt{y} are created at this
stage. These fragments are then composed to form program structures
(some examples are shown as pseudo-C to the right), and finally compiled
to LLVM IR where instructions are added and testing is performed (bottom
box).

\begin{figure*}[t]
  \centering
  \includegraphics[width=\textwidth]{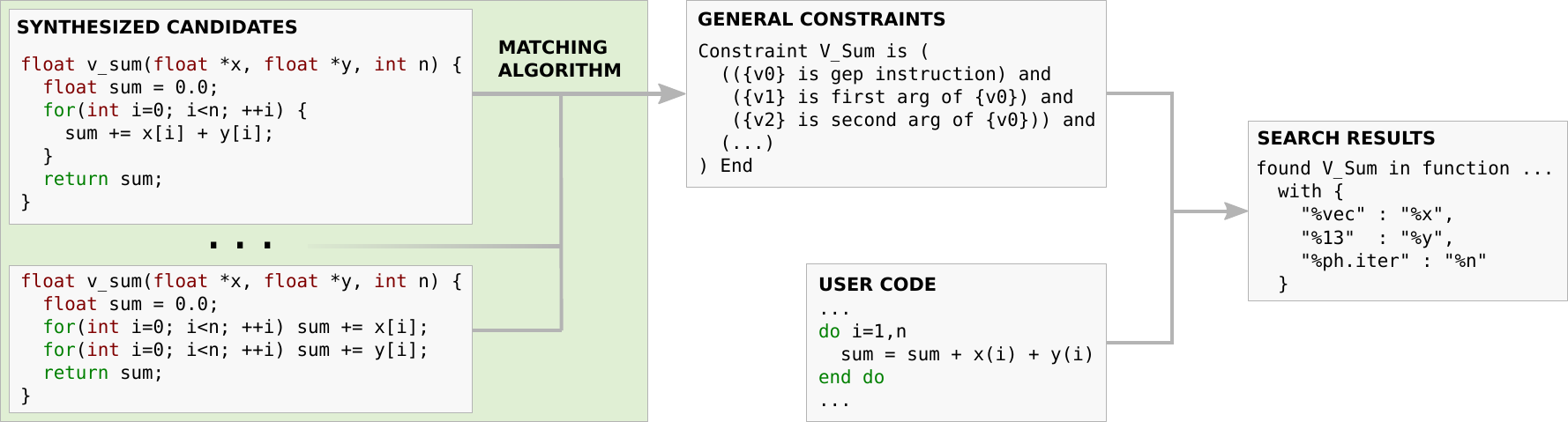}
  \caption{Generating constraints from synthesized candidates:
  Our matching algorithm transforms synthesized programs into constraints and
  then constructs a generalized constraint description in IDL \cite{Ginsbach2018}.
  This searches user code and generates matches.
  Our contribution (generating constraints from examples) is highlighted in
  \textcolor[HTML]{81a663}{green}.}
  \label{fig:fan}
\end{figure*}

\subsection{Generating Programs} \label{sec:dataflow}

Given a control flow structure composed from several fragments, the
final step in our synthesis process is to add instructions to it,
producing a candidate program. We use a generic algorithm to do this,
with no knowledge of the specific fragments that comprise the control
flow structure. We walk the dominance tree of the control flow graph,
inserting stochastically sampled instructions in the appropriate places.
Additionally, $\phi$ nodes are used to handle looping or divergent
control correctly. This algorithm is shown in \Cref{alg:dataflow}.

At each node in the dominance tree, a set of `live' instructions is
maintained.  The initial instructions in this set come from fragment
specifications (for example, values loaded from memory in a loop
iteration). New instructions are sampled from a set of possibilities: we
allow for integer and floating point arithmetic, calls to intrinsic
mathematical functions, and a small number of other simple primitives
such as conditional selects. Searching for a correct candidate program
amounts to iteratively performing this instruction generation algorithm
on each possible control flow structure, testing each resulting program
for behavioral equivalence until a solution is found.

By exploiting annotated type signatures and general heuristics, we can
realistically synthesize complex imperative programs. \Cref{sec:synth}
discusses the complexity and time requirements of this process.

\subsection{Behavioral Equivalence}\label{sec:behavioral}

Our methodology is concerned with `black-box' interfaces.  We are able
to observe input-output behavior, but have no ground truth to evaluate
the correctness of potential solutions against.  We judge a solution as
behaviorally equivalent to a reference implementation, if it behaves
equivalently across a large set of random example inputs.  More
formally, for a solution $s$, reference $r$ and input vectors $
\mathbf{x}_i $.
\[ correct(s) \iff \forall i \; . \; s(\mathbf{x}_i) \approx (\mathbf{x}_i) \]
An approximate notion of equality ($\approx$) is used to compare
floating point values.  This correctness decision is unsound; there is
no way to establish formally that a candidate program will behave
correctly on every possible input.  However, work on property-based
testing \cite{Claessen2000} uses a similar assumption to ours---if
enough is known about the way inputs are used by a program, then
equivalence over a large number of random samples strongly implies
equality.  In practice, users could be asked to sign off on  synthesized
programs.

\section{Recognizing Learned Programs}
\label{sec:conversion}
In \Cref{sec:method} we described a technique for synthesizing LLVM programs
based on input-output examples and information provided by library vendors about
their function call interfaces. This allows us to model existing and new target
library interfaces.

We use these generated LLVM programs to develop a constraint description
of the underlying library implementation. These constraints are then
used to automatically detect relevant sections in existing code bases.
Searching for exact matches of synthesized LLVM IR fragments is unlikely
to achieve any success.  The synthesized code itself is not a sufficient
model, as the IR is not normalized and there are many potential
alternative implementations. In order to generate a useful model, we
need to extract common features from multiple implementations.

In this section we will establish an approach to converting sets of
synthesized IR fragments into constraint programs as shown in
\Cref{fig:fan}.  The contribution of this paper is highlighted in the
green section of the figure: synthesized programs are merged with a
graph matching algorithm and a constraint description is generated from
the result.  This constraint description can then be applied to user
code in order to detect matching instances.

Firstly, we will show how individual fragments can be converted into
candidates.  Secondly, we  show how multiple fragments can be
accumulated, crystallizing out the most important features using graph
constraint matching.  Thirdly, we will show how these accumulated
fragments can be converted into constraints which are used to search the
developer's program.

\subsection{Generating Constraints From LLVM IR Fragments}

To build a constraint description of an LLVM IR fragment, we abstract
the SSA-form code into a graph $G = (V, E)$ with edges $ E \subset
\mathbb{N}\times V\times V$.

\begin{figure*}[t]
  \centering
  \includegraphics[width=\textwidth]{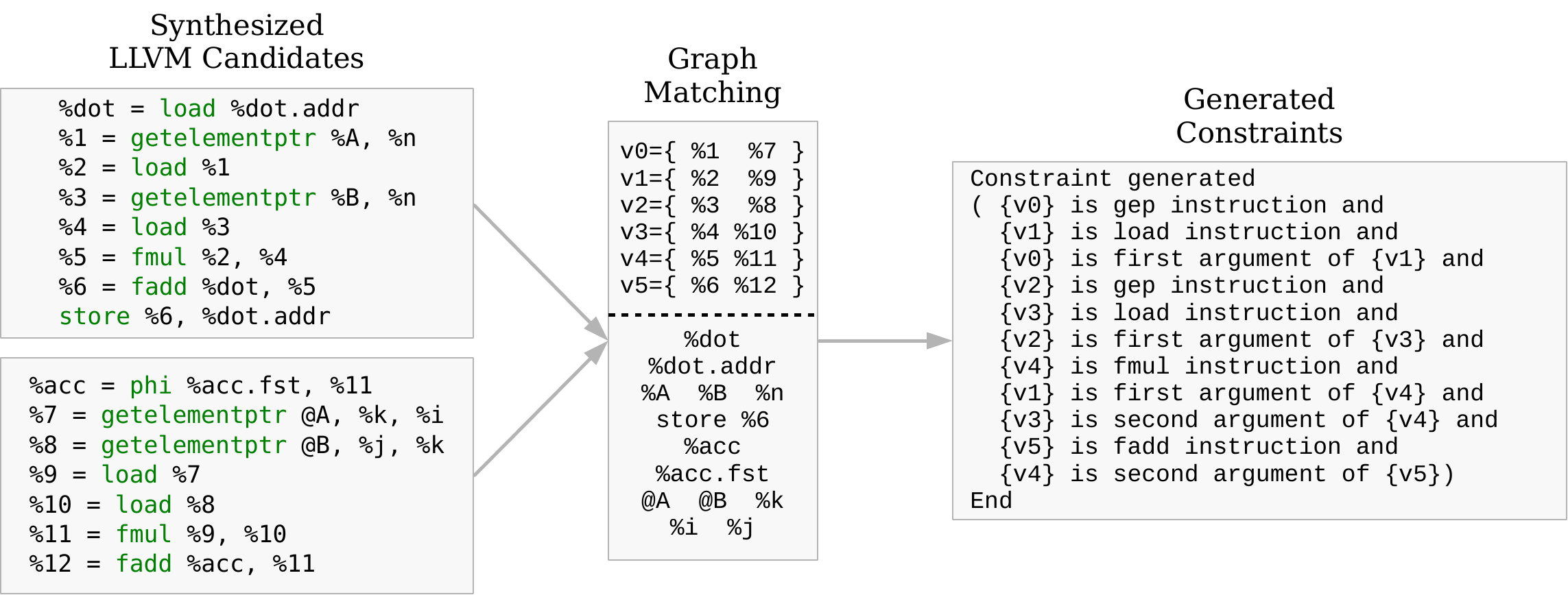}
  \caption{Generalizing Constraints: 
  Candidate constraints from two LLVM IR versions of a dot product are matched.
  The algorithm identifies six groups of instructions to be matched together
  as shown in the top half of the middle graph matching box.
  The others are discarded as shown in the lower half.
  The generalized constraints are then output as shown in the final box.}
  \label{fig:matching}
\end{figure*}

We use the notation $a\xrightarrow{n}b$ to signify $(n,a,b)\in E$, which
means that instruction $a$ is the $n$th argument of instruction $b$.
Function parameter and constants are modeled as instructions with
special opcodes.  We can now mechanically generate a constraint
description that specifies the exact combination of instructions and
argument relations from the program.  In this specific example, the
output is in the IDL programming language \cite{Ginsbach2018}
(\textsc{general constraints} in \Cref{fig:matching}).

This results in a constraint program that detects exact matches.
However, we want to capture a wider class of possible functionally
equivalent implementations rather than a single exact program. In the
next section we describe a graph-based method for generalizing these
constraints.

\subsection{Matching LLVM IR Fragments Together}

In order to generate a common constraint description of a representative
set of LLVM IR functions, we need to match together the respective graph
representations. The aim of this is to uncover the essential parts of
the computation. These are then extracted and turned into constraints.

Given a set of LLVM IR programs, we construct a single graph by directly
combining their respective vertices and edges. We now aim to match together
nodes from the different programs that are semantically equivalent. For this, we
will derive an equivalence relation $\sim$ that partitions the vertices into a
set of equivalence classes $V/\sim$. Vertices that correspond to the same
behavior across examples will belong to the same class. We use the usual
notation for the equivalence class $\bar{u}$ of a vertex $u$: $ \bar u:=\{v\in
  V\mid u\sim v\} $, and write $\bar x\xrightarrow{n}\bar y$ to express the
  following properties:
\begin{align*}
    \forall a \in \bar{x} \;.\; \exists b \in \bar{y} \;.\; a \xrightarrow{n} b \\
    \forall b \in \bar{x} \;.\; \exists a \in \bar{y} \;.\; a \xrightarrow{n} b
\end{align*}

These properties ensure that each instruction in class $\bar x$ is the
$n$th argument of an instruction in $\bar y$, and for each instruction
in $\bar y$, the $n$th argument is in $\bar x$.

\subsubsection{Deriving An Optimization Target}

We intuitively require that an approximation of several properties
should hold. Firstly, instructions that are mapped onto each other
should have the same opcode:
\begin{align*}
    a\sim b\implies op(a)=op(b)
\end{align*}
Secondly, edge relationships should be compatible with the equivalence
relation:
\begin{align*}
    x\xrightarrow{n} y\implies\bar x\xrightarrow{n}\bar y
\end{align*}
Thirdly, we do not want argument relationships to be collapsed together by the
equivalence relation:
\begin{align*}
a\xrightarrow{n}b\implies&\bar a\neq\bar b\\
a\xrightarrow{n}c\land b\xrightarrow{m}c\land \bar a =\bar b\implies&a=b\lor n=m
\end{align*}

Of course, we cannot expect all the criteria to be perfectly fulfilled.
Instead we aim for an approximate result. We thus define a metric $m$
on the set of possible equivalence relations $EQ(V)$ that punishes any
deviation from the previously given conditions.
\begin{align*}
m(\sim)=(p_1\cdot&\left|V/\sim\right|\\
        +p_2\cdot&\left|\{ v\in V/\sim\mid\exists x,y\in \bar v\colon op(x)\neq op(y)\}\right|\\
       +p_3\cdot&\left|\{\bar v\in V/\sim\mid \exists y\in \bar v,x,n\colon x\xrightarrow{n}y\land \neg \bar x\xrightarrow{n}\bar y)\}\right|\\
       +p_4\cdot&\left|\{\bar v\in V/\sim\mid\text{third conditions not satisfied}\}\right|)^{-p_5}
\end{align*}
The first parameter controls how much the equivalence relation is
encouraged to merge together vertices into larger equivalence classes.
The next three parameters control how strictly each property is
enforced. The fifth parameter is simply used to change the distribution
of the resulting scores without changing their relative order.

By trial and error we assigned the following values to each parameter:
$p_1=1.0$, $p_2=0.5$, $p_3=0.5$, $p_4=0.5$.

\subsubsection{Optimizing the Metric}

Having defined the metric, we now need to find the optimal solution,
i.e.\ the partition of the graph that maximizes the metric. We use a
simple evolutionary algorithm over 1000 generations to maximize $m$.

\begin{table*}[t]
  \begin{minipage}{0.49\linewidth}
    \normalsize
    \centering
    \begin{tabularx}{\columnwidth}{lllX}
      \toprule
      \textbf{Name} & \textbf{Kernels} & \textbf{Acceleration} & \textbf{LoC}  \\
      \midrule
      NWChem & Dense  & BLAS &      1.2M\\
      Abinit & Dense  & BLAS,CUDA & 900k\\
      Pathsample & Sparse  & Handcoded SpMV& 40k \\
      Darknet & Neural Network & CUDA& 27k  \\
      Parboil & Linear Algebra & Handcoded MxM &  187\\
      \bottomrule
    \end{tabularx}
    \caption{Application source code used for evaluation.}
    \label{tab:sources}
  \end{minipage}\hfill
  \begin{minipage}{0.49\linewidth}
    \normalsize
    \centering
    \begin{tabularx}{\columnwidth}{llX}
      \toprule
      \textbf{Library} & \textbf{Platform} & \textbf{Kernels} \\
      \midrule
      Intel MKL   & Intel CPU       & Dense Linear Algebra \\
      cuBLAS      & Nvidia GPU      & Linear Algebra \\
      cuDNN       & Nvidia GPU      & Neural Networks \\
      cuSparse    & Nvidia GPU      & Sparse Linear Algebra \\
      CLBlast     & OpenCL Devices  & Dense Linear Algebra \\
      \bottomrule
    \end{tabularx}
    \caption{Optimized libraries selected for evaluation.}
    \label{tab:libraries}
  \end{minipage}
\end{table*}

\subsection{Generating Constraints from a Matching}

Given an appropriate equivalence relation, we can emit a constraint
program in a straightforward fashion.  To do this, we firstly generate a
new graph $G/\sim:=(V/\sim,E/\sim)$.  Here, the set $E/\sim\subset
\mathbb{N}\times V/\sim\times V/\sim$ is defined by the following
property, where $ (n, \bar a, \bar b) \in E/\sim $ iff $ \bar a
\xrightarrow{n}\bar b $.

We then define a threshold $d$ and remove all vertices of the graph
$G/\sim$ that contain fewer than $d$ elements.  This results in the
removal of specific quirks of the individual synthesized programs,
leaving the essential algorithmic skeleton intact. For the value of
$d$, we choose the number of merged programs.

\subsubsection{Example}

Consider the example in \Cref{fig:matching}. On the left, we can see two
simplified pieces of LLVM IR code from different loops.  The first is
from the body of a loop that computes the dot product of two vectors
given as pointers, the second is from the innermost loop of a naive
matrix multiplication kernel.  They are matched together with the
introduced graph matching algorithm, resulting in the clustering that is
displayed in the middle column of the figure.

After discarding all equivalence classes with fewer than two elements, we
generate constraints mechanically.  Features that were specific to one program
are discarded, notably the instructions \texttt{\%dot}, \texttt{\%acc},
\texttt{\%12} and the store instruction as seen in \Cref{fig:matching}. We can
now use the generated constraints to find equivalent code sections in other
source code \cite{Ginsbach2018}.

\subsection{Suggesting Replacements}

Given a set of generated constraints as described above, the final step in our
process is to search through user code for matching instances. The SMT-based
search algorithm is described in detail by previous work
\cite{Ginsbach2018,Ginsbach2018a}; we treat this process as a black box.

We first  ensure that the generalized constraints include a mapping for
each function argument (if this is not the case, we discard the
constraints). From a search result, we can then identify which matched
value corresponds to each function argument. Previous work targeting
precisely known library functions \cite{Ginsbach2018} makes an automated
replacement at this point by operating at the compiler IR level.
However, as we require the developer to ``sign off'' on a replacement,
we map the IR values back to source locations. We do this by exploiting
Clang SSA value naming with the \texttt{-fno-discard-value-names} flag,
as well as line and character debug information. This allows us to
identify replacement opportunities.

\begin{figure*}
  \centering
  \includegraphics[width=\textwidth,height=0.4\textwidth]{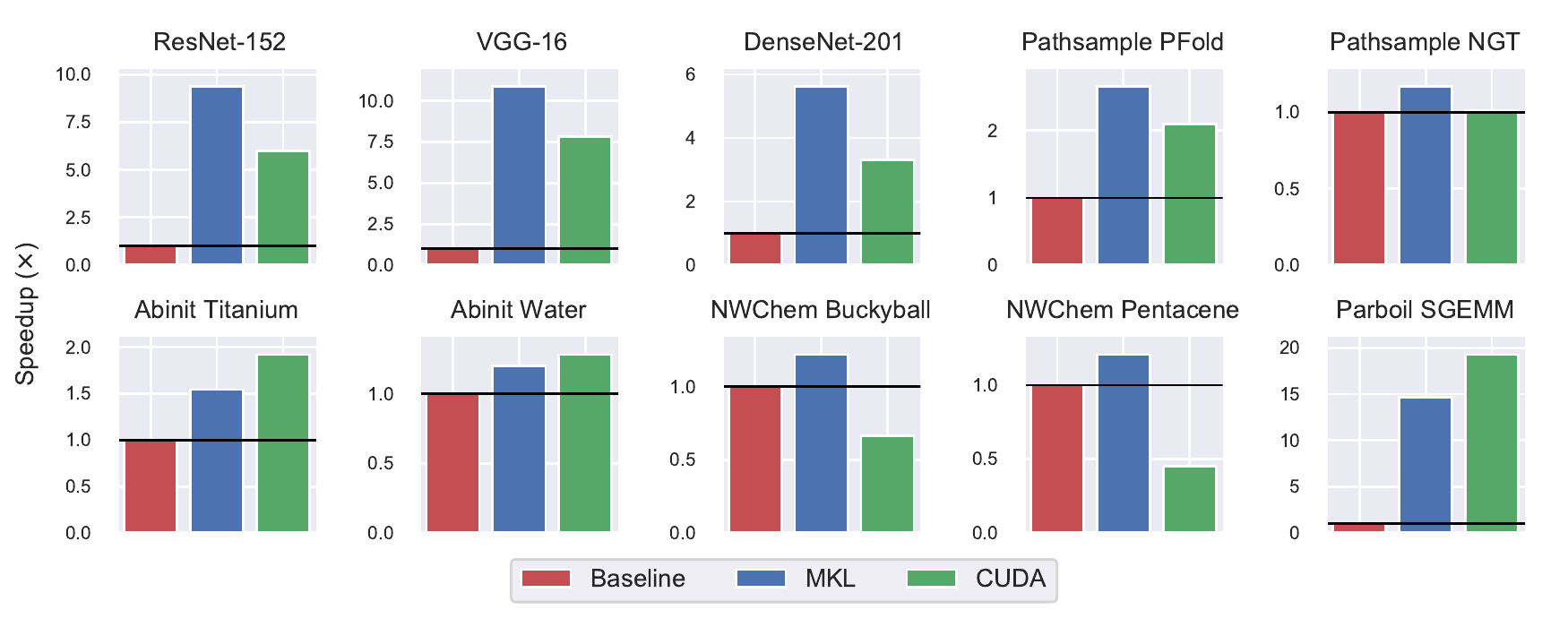}
  \caption{Performance achievable by adopting code replacements suggested by our
    tools, for both Intel MKL and Nvidia CUDA libraries across the set of
  benchmarks listed.}
  \label{fig:perf-results}
\end{figure*}

\section{Setup}
\label{sec:setup}
\subsection{Applications}

We selected compute intense applications that are likely to benefit from
acceleration libraries. These are shown in \Cref{tab:sources}. Three of these
are scientific applications: NWChem, Abinit and Pathsample, each representing a
large code base.  hese are significant applications; Abinit alone has been
cited more than 6,000 times since 2002.

Abinit must be linked to a BLAS implementation installed somewhere on
the target system, while NWChem uses an internal library and Pathsample
implements a small set of required operations by hand (including some
sparse methods).  For each scientific application, we evaluated  two
different standard data sets which correspond to different chemistry
simulations. These data sets were selected to exercise different
portions of the underlying code base.

Neural networks also make large use of acceleration libraries. We
examined Darknet \cite{darknet13}, a widely used, open-source deep
learning framework. It has been used recently to implement a number of
highly cited, state-of-the-art models
\cite{Rastegari2016,Redmon2016,Redmon2017}. The framework consists of
two distinct implementations (in C and CUDA) We evaluated three well
known ImageNet models implemented using Darknet.

Finally, we examined a well known benchmark program from Parboil, SGEMM,
which contains a hand-coded matrix-multiplication. As this is dominated
by one linear algebra routine, it gives an upper bound on the typical
performance achievable by our approach.

\subsection{Libraries}

The libraries (see \Cref{tab:libraries}) fall broadly into two
categories: those that are optimized for a particular CPU architecture
(Intel MKL)  to achieve performance, and those that use the GPU (CUDA
libraries, CLBlast). There are a number of different CUDA libraries that
can be run on NVidia GPUs; for brevity we refer to these together as a
single collection. 

\subsection{Platform}

We targeted an Intel Xeon E5-2620 processor with 24 cores, 16GB of RAM
and an NVidia Tesla K20 GPU. Applications were compiled at \texttt{-O3}.
For the cross platform evaluation we targeted a 12 core AMD A10-7859k
with an integrated AMD Radeon R7 iGPU.

\subsection{Methodology}

We ran each application from its ``out of the box''
configuration on the Intel platform to give a performance baseline.

For Pathsample the baseline code is sequential, handwritten Fortran with no
library calls. This is also the case for NWChem which contains sequential C
(which has hand inlined specimen BLAS implementations) In the case of Abinit,
the baseline links to standard BLAS libraries. In the case of Darknet, there are
2 baselines available and we use the default sequential C baseline.  We
evaluated our performance when targeting  both Intel MKL 2019 and a range of
CUDA 8.0 based libraries.

To evaluate the impact of moving to a new platform, we focused on
Darknet and evaluated our approach on an AMD platform that does not
support CUDA. Instead it targets the CLBlast library.

\section{Results}
\label{sec:results}
We first evaluate the performance of our approach across applications, libraries
and platforms. Next, we examine the number of library calls and candidate
matches for API migration. This is followed by an evaluation of the execution
time needed to synthesize equivalent programs for each library. Finally, we
evaluate the accuracy of our graph matching algorithm and discuss the potential
for unsound behavior to arise when using these tools.

\subsection{Overall Results} \label{sec:perf}

A summary of our performance results is shown in \Cref{fig:perf-results}.  On
scientific applications, we found that the best implementation for each one
achieved speedups of between $1.2$ and $2.7\times$.  This is the end to end
performance of each application rather than just isolated kernels.  In the case
of Pathsample, the NGT configuration spends less time in sparse matrix
operations than the PFold configuration. Amdahl's law means that inevitably
PFold will benefit more from acceleration. MKL outperforms the Nvidia libraries
by a small margin in both cases. If we only used Nvidia libraries, there would
be speedup available in the case of NGT.

This pattern continues with NWChem where MKL significantly outperforms the
Nvidia libraries. Modest speedups are available for both configurations with an
end-end speedup of $1.2\times$ Abinit shows a different behavior, where the Nvidia
libraries outperform MKL, giving $1.2$ to $1.9\times$ speedup. This is possibly due
to the increased array sizes where the benefits of acceleration outweigh
communication overhead. Unlike NWChem, both acceleration libraries improve
performance.

We see more significant improvement for the DNNs as the amount of time spent in
accelerator code sections is much greater. Improvements range from $5.5\times$
for the smaller DenseNet-201 to $11\times$ for the largest network: VGG-16. Like
Pathsample and NWChem, all the the DNNs achieve the greatest performance with
MKL, though Nvidia libraries still give improvements: $3.2\times$ to
$7.7\times$.  The impact of Amdahl's law can be clearly seen for Parboil SGEMM.
Here there is just one kernel that can be readily accelerated. It achieves
$15\times$ to $19\times$ speedups and provides a best case example.

\begin{table*}[t]
  \centering
  \scriptsize
  \begin{tabular}{cc|ccccccccccc}
    && \textsc{spmv} & \textsc{gemm} & \textsc{gemv} & \textsc{ger} &
    \textsc{axpy} & \textsc{axpby} & \textsc{scal} & \textsc{copy} &
    \textsc{dot} & \textsc{softmax} & \textsc{relu} \\
    \midrule
    \multirow{4}{*}{Abinit} 
    & \textbf{P} & & 180 (180) & 47 (47) & & 21 (21) & 2 (2) & 20 (20) & 70 (70) \\
    & \textbf{TP} & & 0/0/180/180 & 0/0/47/47 & & 21/21/21/21 & 0/2/2/2 & 20/20/20/20 & 70/70/70/70 \\
    & \textbf{FP} & \\
    & \textbf{FN} & & 180/180/0/0  & 47/47/0/0 & & & 2/0/0/0 \\
    \midrule
    \multirow{4}{*}{Pathsample} 
    & \textbf{P} & 2 (0) & 1 (0) & 1 (0) & 3 (0) & 7 (0) & & 13 (0) & 5 (0) & 1 (0) \\
    & \textbf{TP} & 0/0/2/2 & 0/0/1/1 & 0/0/1/1 & 3/3/3/3 & 7/7/7/7 & & 13/13/13/13 & 5/5/5/5 & 1/1/1/1 \\
    & \textbf{FP} & \\
    & \textbf{FN} & 2/2/0/0 & 1/1/0/0 & 1/1/0/0 \\
    \midrule
    \multirow{4}{*}{NWChem} 
    & \textbf{P} & 2 (0) & 2 (0) & 2 (0) & 2 (0) & 2 (0) & 27 (0) & 2 (0) & 2 (0) & 2 (0) \\
    & \textbf{TP} & 0/0/2/2 & 0/0/2/2 & 0/0/2/2 & 0/2/2/2 & 0/2/2/2 &
    0/27/27/27 & 0/2/2/2 & 0/2/2/2 & 0/2/2/2 \\
    & \textbf{FP} & & & & & & & 0/2/2/0 & 0/5/5/0 \\
    & \textbf{FN} & 2/2/0/0 & 2/2/0/0 & 2/2/0/0 & 2/0/0/0 & 2/0/0/0 &
    27/0/0/0 & 2/0/0/0 & 2/0/0/0 & 2/0/0/0 \\
    \midrule
    \multirow{4}{*}{Darknet}
    & \textbf{P} & & 2 (1) & 1 (0) & & 1 (0) & & 1 (0) & 1 (0) & 1 (0) & 1 (0) & 1 (0) \\ 
    & \textbf{TP} & & 0/0/2/2 & 0/0/1/1 & & 0/1/1/1 & & 0/1/1/1 & 0/1/1/1 & 0/1/1/1 & 0/0/0/0 & 0/1/1/1 \\
    & \textbf{FP} & & & & & & & 0/3/3/0 & 0/2/2/0 & 0/1/1/0 \\
    & \textbf{FN} & & 2/2/0/0 & 1/1/0/0 & & 1/0/0/0 & & 1/0/0/0 & 1/0/0/0 & 1/0/0/0 & 1/1/1/1 & 1/0/0/0 \\
    \midrule
    \multirow{4}{*}{Parboil}
    & \textbf{P} & & 1 (0) \\ 
    & \textbf{TP} & & 0/1/1/1 \\
    & \textbf{FP} \\
    & \textbf{FN} & & 1/0/0/0 \\
  \bottomrule
  \end{tabular}
  \caption{Instances of each function category discovered across the
    different applications evaluated. The first row for each application
    gives the total number of potential matches we identified by
    hand-examination (positives, \textbf{P}), with the number of these
    corresponding to inlined library calls given in parentheses.
    Subsequent rows give the number of correctly identified
    opportunities (true positives, \textbf{TP}), incorrect matches
    (false positives, \textbf{FP}) and missed opportunities (false
    negatives, \textbf{FN}). Results are quoted as $x/y/z/w$ for the
    four different versions of our discovery algorithm: no
  generalization, generalization, generalization with nested loop
corrections, and false positive testing.  See \Cref{sec:detection} for
details.}
  \label{tab:detectstats}
\end{table*}

\subsection{Porting to New Hardware} \label{sec:port}

\begin{figure}[t]
  \centering
  \includegraphics[width=\columnwidth,height=0.35\columnwidth]{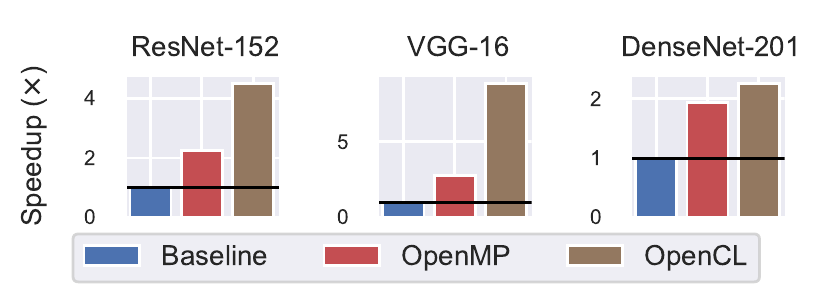}
  \caption{Performance results for neural network inference on an AMD device
  with no CUDA support.}
  \label{fig:nnperf}
\end{figure}

Within Darknet, the use of optimized GPU libraries is built into the code: CUDA
and CPU implementations are mixed together using preprocessor directives and the
build system.  As CUDA is not available on AMD GPU platforms, porting Darknet to
such a platform means targeting OpenCL based libraries such as the CLBlast
library \cite{Nugteren2018}, the results of which are shown in
\Cref{fig:nnperf}.

We compared the performance of ``out-of-the-box'' Darknet against a handwritten
OpenMP version \cite{8573503}, and our approach. The results of this comparison
are shown in \Cref{fig:nnperf}. On all three networks, our approach outperforms
the OpenMP implementation which represents the best readily-available CPU
performance on an AMD processor. We achieve speedups from $2.4\times$ on DenseNet-201
to $9\times$ on VGG-16.  DenseNet-201 performs smaller matrix multiplications
than the other networks, and so benefits less from GPU execution. Our results
show that our approach allows for programmers to port applications to other
platforms, without having to support multiple code bases for each possible
implementation.

\subsection{Library API usage} \label{sec:discovery}

\Cref{tab:detectstats} shows the number of library call sites we detected in the
original applications. For simplicity. we group functions that perform the same
abstract computation together.  For example, \texttt{cublas\_sgemm},
\texttt{cblas\_sgemm} and \texttt{clblast::Gemm<float>} are all considered
together in the \textsc{gemm} group.

Some of the applications we examined make extensive use of library functions.
For example, Abinit links against an installed standard BLAS library, and so all
the instances we detect in its code are from inlined library calls. Other
applications bundle their own implementations; our approach detects this code
rather than the corresponding call sites which results in a smaller overall
number of matches. The true positive figures (\textbf{P}) in
\Cref{tab:detectstats} show both the total number of potential matches and the
number that come from inlined library calls.

\subsection{Synthesis} \label{sec:synth}

The time taken for our synthesizer to correctly synthesize each library program
is acceptable for our usage model: every example could be synthesized in under 4
hours on a desktop-class machine, with examples that use shorter instruction
sequences taking far less time. Synthesis time was not a primary goal of our
work---learning the behavior of a function is a one-off task.  If synthesis time
is a bottleneck, there are many existing approaches to improve performance
\cite{Gulwani2011a,Schkufza2013}, but this is beyond the scope of this paper.

\subsection{Matching} \label{sec:detection} 

\Cref{tab:detectstats} shows the results obtained when searching for code
satisfying our generated constraints. We tested four different versions of the
constraints: those generated from a single example constraint-based program,
generalized versions from multiple programs, generalized with a post-processing
step, and finally with dynamic testing of replacements.

The constraints generated from a single program fail to discover many examples.
Only simple, inlined library calls are consistently matched by these constraints
(Abinit in \Cref{tab:detectstats}, as the inlined code is identical to the code
from which constraints are generated.  

We then applied our graph matching algorithm to generalize constraints. These
constraints are more successful; many instances that were not previously matched
now are (e.g.\ in Darknet). Some instances such as \textsc{gemm} and
\textsc{spmv} were not discovered by the generalized constraints. We discovered
this was due to a consistent difference between Clang's code generator and the
synthesizer for nested loops---a mechanical post-processing step fixed these
constraints, allowing these examples to be detected properly (\textsc{gemm},
\textsc{gemv}, \textsc{spmv} columns in \Cref{tab:detectstats}).

Although these constraints generalized well, some false positive matches occur
due to over-generalization (e.g.\ for \textsc{scal}, \textsc{copy} in NWChem and
Darknet). We found that this was due to missing data dependencies in code that
interleaved another task with the learned function. To address this, we
performed dynamic testing of suggested replacements using IO examples,
eliminating all false positives we observed.

The only example not to be discovered in any of our test codes was
\textsc{softmax}: it was implemented in the code we examined using a common
numerical trick where the input data is shifted uniformly by its maximum value.
The synthesizer is not able to learn this approach.  Fortunately, it was not a
significant contribution to execution time in the programs examined.

\subsection{Soundness}

There are a number of ways in which unsound behavior can arise when
using our synthesis, generalization and replacement suggestion tools:
\begin{itemize}
  \item Random IO examples may not capture the full range of a
    function's behavior. While this is a theoretical limitation of our
    synthesizer, in practice we have not encountered any function that
    suffers from this problem.
  \item The synthesizer may fail to synthesize a library function at
    all: not all functions have behavior that can be captured by the
    fragments used in this paper. If this happens, the function is
    ignored. Our technique demonstrates useful performance improvements
    despite not being able to learn every library function.
  \item False positives and negatives can occur when matching
    constraints. We found that our constraints generalized well to
    detect complex examples, and that false positives can be readily
    eliminated by dynamic testing.
\end{itemize}

While these sources of unsound behavior can and do affect our process in
some cases, the actual effects are not critical to the practical
application of our tools.

\subsection{Scalability}

A natural question to be asked of our tooling is how well it scales beyond the
examples shown in our existing evaluation. For example, in the context of
machine learning we may wish to synthesize the behavior of a batch normalization
layer or pooling. We anticipate that our approach will scale well to problems
such as these. Other, more different problem domains such as sorting algorithms
(a topic of interest in the program synthesis literature \cite{Rosin2018}) could
be synthesized by expanding the set of fragments.

\section{Related Work} \label{sec:related}
\noindent {\bf Rejuvenating Code: }
Most approaches within the software engineering community use a form of
user guided refactoring \cite{dig2005role} to perform simple syntactic
restructuring of application code, supported by standard IDEs such as
Eclipse \cite{murphy2006java}.  In \cite{Catchup}, replaying of
refactoring for changed APIs calls is presented.  However, refactoring
techniques assume complete knowledge of library behavior.  Such
approaches do not address matching existing user code to emerging
library APIs \cite{brito2018apidiff}.

\noindent {\bf Automatic Acceleration: }
Our work uses the IDL constraint language \cite{Ginsbach2018} as a way
to express computational patterns shared by a set of synthesized
programs, and we make use of the detection features of IDL to discover
instances of user code that match these patterns.  While we
automatically generate constraint descriptions for library interfaces
based on their synthesized behavior, \cite{Ginsbach2018} requires an
expert compiler developer to write constraint descriptions by hand;
they do not consider
synthesizing examples to drive constraint generation nor the
generalization of constraints.

\noindent {\bf Sketching:}
Our technique uses control flow fragments to express partial guesses of
the structure of a possible solution.  \cite{Solar-Lezama2009}
introduces the idea of \emph{sketching} to allow the programmer to
express a partial solution to a synthesis problem. Recent work in this
area deals with recursive tree transformations \cite{Inala2017} and
modularity of sketches \cite{Singh2014}.

Our work lies within the area of \emph{sketch generation}, where the
synthesis problem is split into two parts: the first where sketches are
generated, and the second where they are instantiated to produce
solutions. Our approach lies within this space. \textsc{Scythe}
\cite{Wang2017} uses this technique to generate SQL queries, and
the \textsc{LaSy} language \cite{Perelman2014} uses libraries
of composable domain-specific functions to describe the space of
possible solutions, but requires a carefully-chosen set of input-output
examples to work effectively.

\noindent {\bf Synthesizing Imperative Programs:} 
Program synthesis often deals with synthesis of highly composable
functional programs, which often allows synthesized programs to be
represented in a minimal normal form \cite{Feser2015}. Our technique
generates LLVM intermediate representation, which does not permit a
`normal form' in the same way.  Other work in which imperative programs
are synthesized often deals only with straight line code
\cite{Gulwani2011a, Sasnauskas2017}, or treats imperative control flow
as special-cased components \cite{Gulwani2011}.  More commonly targeted
than LLVM IR is assembly code, especially for superoptimization:
\cite{Srinivasan2015, Schkufza2013, Heule2016, Hasabnis2015b}. Our
combination of target representation and treatment of control flow is a
novel one.

\noindent {\bf Synthesizing for Acceleration:}
Other work has used program synthesis as a mechanism by which programs
can be automatically accelerated.  For example, recent work uses program
synthesis to generate parallel versions of sequential code, with a focus
on numerical array-processing algorithms with single-pass control flow
\cite{Farzan2017a,Fedyukovich2017a}.  The space of programs tackled
however is highly restricted.  Helium \cite{mendis:pldi:2015} uses
syntax-guided synthesis to synthesize programs in the Halide
\cite{Ragan-Kelley2013} image processing DSL. This approach has
explicit knowledge of stencils hardwired into the synthesis phase and
cannot be applied to unknown libraries. 

Similarly, in \cite{Kamil2016a}, loop verification conditions are
extracted from Fortran programs using inductive program synthesis
techniques. These conditions can then be translated mechanically to
Halide. Our approach to program synthesis is broader in application at
the expense of formal verification.

\noindent {\bf Type-Directed Synthesis}
Our annotated type signatures are similar in spirit to
\emph{type-directed synthesis}.  \textsc{Myth} \cite{Osera2015a} uses
type signatures alongside examples to synthesize recursive functional
programs. More similar to our approach is the idea of using refinement
types to guide the search process \cite{Polikarpova2016}.

\section{Conclusion and Future Work} \label{sec:conclusion}
Porting existing code to exploit accelerator libraries is a challenging
problem for programmers.  Understanding the behavior of existing and new
libraries requires significant work on the programmer's part.

This paper presents two main contributions to help with this API
evolution: a program synthesis technique that uses vendor-supplied type
annotations to infer partial control flow structure for potential
solutions, and a novel graph-matching based approach to finding a common
description for a set of input programs. Using this approach we were
able to achieve significant improvements to existing large scale code
bases.

While our approach uses code normalization to aid matching, future work
will focus on improving program synthesis to more closely match user
code. This can be achieved by adding priors over the synthesis search
space to bias construction.  Currently the graph-matching algorithm
introduces false positives that are eliminated with dynamic information.
Future work will investigate the use of iteratively generating programs
from generalized constraints and testing for equivalence against
synthesized examples using SMT solvers, eliminating false positives.

\bibliographystyle{IEEEtran}
\bibliography{references}

\end{document}